\documentclass[12pt,preprint]{aastex}

\newcommand{\myemail}{lic@nju.edu.cn}

\shorttitle{Proton activity of solar cycle 24}
\shortauthors{Li, Miroshnichenko, and Fang}

\begin{document}


\title{Proton activity of the Sun in current solar cycle 24}


\author{C. Li\altaffilmark{1,2}, L. I. Miroshnichenko\altaffilmark{3,4}, and C. Fang\altaffilmark{1,2}}
\altaffiltext{1}{School of Astronomy and Space Science, Nanjing University, Nanjing 210093, China. \myemail}
\altaffiltext{2}{Key Laboratory for Modern Astronomy and Astrophysics (Nanjing University), Ministry of Education, Nanjing 210093, China}
\altaffiltext{3}{N. V. Pushkov Institute of Terrestrial Magnetism, Ionosphere and Radio Wave Propagation (IZMIRAN),Russian Academy of Sciences, Troitsk, 142190 Moscow Region, Russia}
\altaffiltext{4}{D. V. Skobeltsyn Institute of Nuclear Physics (SINP), M. V. Lomonosov Moscow State University, 1(2), Vorobyevy Gory, 119234 Moscow, Russia}


\begin{abstract}
We present a study of 7 large solar proton events (SPEs) of current solar cycle 24 (from 2009 January up to date). They were recorded by GOES spacecraft with highest proton fluxes over 200 pfu for energies $>$10 MeV. In situ particle measurements show that: (1) The profiles of the proton fluxes are highly dependent of the locations of their solar sources, namely flares or coronal mass ejections (CMEs); (2) The solar particle release (SPR) times fall in the decay phase of the flare emission, and are in accordance with the times when the CMEs travel to an average height of 7.9 solar radii; (3) The time differences between the SPR and the flare peak are also dependent of the locations of the solar active regions (ARs). The results tend to support the concept of proton acceleration by the CME-driven shock, even though there exists a possibility of particle acceleration at flare site with subsequent perpendicular diffusion of accelerated particles in the interplanetary magnetic field (IMF). We derive the integral time-of-maximum (TOM) spectra of solar protons in two forms: a single power-law distribution and a power law broken with an exponential tail. It is found that the unique Ground Level Enhancement (GLE) event on 2012 May 17 displays a hardest spectrum and a largest broken energy that may explain why the this event could extend to relativistic energy.
\end{abstract}


\section{Introduction}

Solar energetic particles (SEPs), with energies from a few keV to $\geq$10 GeV, are produced by the rapid release of magnetic energy during solar eruptions. Among them solar protons are one of significant populations of interplanetary particle streams that determine important properties of space weather. At the Earth's orbit these protons are registered as a solar proton event (SPE). Usually, it is defined as a flux of $\geq$10 MeV protons greater than 1 pfu, or particle flux unit (1 pfu = 1 $particle\ cm^{-2}\ s^{-1}\ sr^{-1}$), and just those proton energy-intensity thresholds are applied in research work (e.g., Miroshnichenko 2003, and references therein). On some practical reasons (see discussion in Miroshnichenko 2003), the NOAA Space Environment Service Center (NOAA SESC) many years ago suggested to use an intensity of 10 pfu for $\geq$10 MeV protons as a reliable signature of SPE. In our study of the most intensive SPEs of the solar cycle 24 we use the NOAA SESC criterion. According to the survey of NOAA SESC, a total number of 256 SPEs were recorded by the Geostationary Operational Environment Satellite (GOES) from 1976 January up to date ($http://www.swpc.noaa.gov/ftpdir/indices/SPE.txt$).

SPEs are always associated with flares and coronal mass ejections (CMEs) concomitantly, and both of them are theoretically capable of accelerating charged particles to high energies (Ellison $\&$ Ramaty 1985; Litvinenko 1996; Somov $\&$ Kosugi 1997; Zank et al. 2000). Whether SPEs originate from flaring active regions (ARs) in the solar corona or from CME-driven shocks propagating through corona to interplanetary space remains controversial (Aschwanden 2012; Gopalswamy et al. 2012; Li et al. 2012; Miroshnichenko et al. 2013; Reames 2013). The relativistic energy extension of SPEs, which can generate sufficient secondary particles registered by ground-based neutron monitors (NMs), is the so-called Ground Level Enhancement (GLE) of solar cosmic rays (SCRs). The first GLE event was evidently observed on 1942 February 28 (Forbush 1946). Since then 71 GLE events have been recorded by the worldwide network of cosmic ray stations in the last 7 solar cycles. Understanding where and how particles are accelerated to high energies in SPEs, especially in GLE events, is still one of the main topics in solar physics.

The intensities of non-relativistic particles resulting in GLE events are generally larger than those of normal ones, and the associated flares and CMEs are generally stronger and faster (Kahler et al. 2012; Mewaldt et al. 2012; Shea $\&$ Smart 2012). This is, however, not the case for the recent GLE 71 occurred on 2012 May 17 (see references in Gopalswamy et al. 2013; Li et al. 2013; Papaioannou et al. 2014; Plainaki et al. 2014). In particular, among the 7 most intensive SPEs in this study, the GLE 71 shows a lowest proton peak flux of 255 pfu at energy threshold $\geq$10 MeV, and it was associated with a relatively weak flare (M5.1 class) and slow CME (1582 km/s). With this inspiring insight, we concentrate below on the key aspects that motives this study: (1) general activity weakness of the current cycle; (2) proximity of highest GOES energy channel ($\sim$433 MeV) to the limit cutoff energy (or to cutoff rigidity of 1 GV) for polar NMs; (3) possible registration of ¡°hidden, poor-identified GLEs¡± by polar NMs. In this context, we try to interpret the observational and physical differences between the unique GLE 71 and the other intensive SPEs occurred in current solar cycle up to now.

\section{Observations and data analysis}

A common consensus has been reached that the current solar cycle 24 is still in a very low activity. In particular, since 2009 January up to date only 7 large SPEs occurred with proton peak fluxes over 200 pfu at energies $>$10 MeV, comparing 20 similar intensive SPEs during the same period of previous solar cycle 23. The GLE event on 2012 May 17 is the first and single GLE event of solar cycle 24. In last solar cycle 23 and during the same period, 9 GLE events had been recorded. Table 1 lists the 7 most intensive SPEs of the current solar cycle 24.

\subsection{Proton fluxes}

The in situ proton measurements are obtained from the GOES 13 spacecraft, which is in geostationary orbit above the Pacific Ocean at L = 6.6 where the geomagnetic effects are minor for protons $\geq$10 MeV. We examine the time-intensity profiles of the 7 SPEs at the energy bands of $>$10, $>$30, $>$50, $>$60, and $>$100 MeV. The numbers and dates of the events under study are following: (1) 2012 January 23; (2) 2012 January 27; (3) 2012 March 7; (4) 2012 March 13; (5) 2012 May 17; (6) 2013 May 22; (7) 2014 January 7. As shown in Figure 1, the spiral pattern of the interplanetary magnetic field (IMF) gives rise to the longitude-dependent solar proton profiles. It can be interpreted as proton acceleration by an expanding shock in the IMF (Cane et al. 1988; Reames 1999). Assuming the strongest acceleration occurs near the central nose of the shock and weakest at the flank, an observer at the Earth should see an initial prompt increase when the solar eruption occurs near the well-connected region ($\sim$W60) and a gradual increase in the far eastern side.

\begin{figure}
   \centering
   \includegraphics[width=38cm]{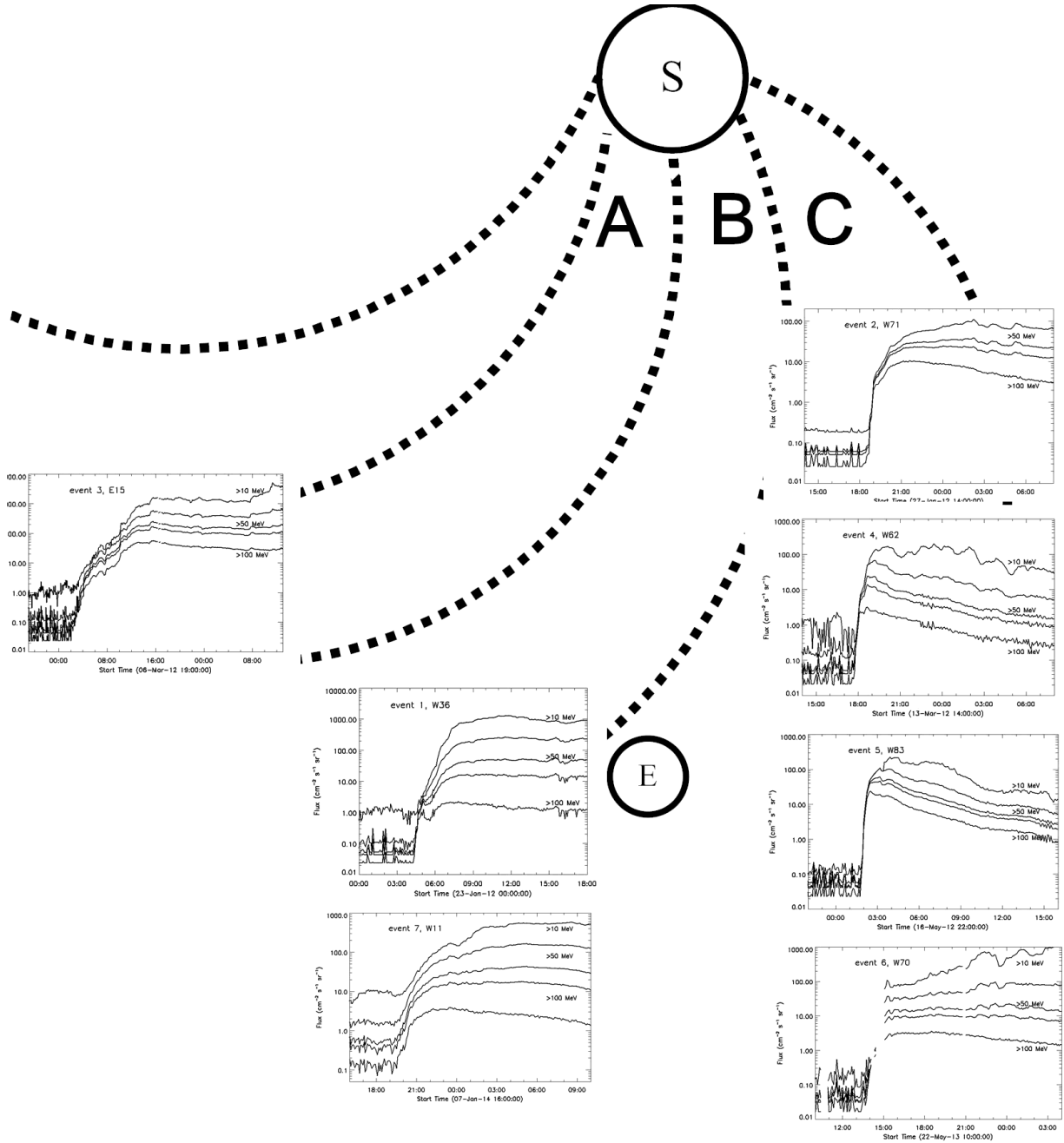}
   \caption{Time-intensity profiles of the 7 SPEs observed by the GOES spacecraft near the Earth. The longitude-dependent profiles are grouped into three types according to the locations of solar sources.}
   \label{fig1}
\end{figure}

The initial time-intensity profiles of the 7 SPEs can be divided into three groups: (1) gradual increase with the solar source located in the eastern hemisphere of region A (event 3, E15), (2) quick increases with the solar sources located not far from the meridian of region B (event 1, W36 and event 7, W11), and (3) very impulsive increases with the solar eruptions occurred near the well-connected region C (event 2,W71, event 4, W62, event 5, W83, and event 6, W70). To be noticed that the time ranges of the profiles of the 7 SPEs in Figure 1 are the same of 18 hours, it makes the comparison of the increase tendencies to be reasonable. The results seem to support the concept that particle are accelerated by an expanding CME-driven shock. However, we can not rule out a possibility of particle acceleration at the flare site with subsequent perpendicular diffusion or cross-field propagation in the interplanetary space (Qin et al. 2011; Giacalone $\&$ Jokipii 2012; Wiedenbeck et al. 2013).

\subsection{Event timing}

The solar particle release (SPR) times with respect to the multi-wavelength observations of solar eruptions is a critical method for determining the particle acceleration source. We first estimate the SPR times of solar protons by subtracting $\Delta t = l/v - 8.3$ minutes from the in situ onset times, where $l$ is the path length of IMF lines deduced from the solar wind model (Parker 1958, 1963), the velocity of solar protons, $v$, is taken to be $\sim$0.7$c$ corresponding to the highest energy channel of 433 MeV for GOES. We then compare the SPR times to the flare soft X-ray (SXR) or hard X-ray (HXR) emission and the CME white-light observation.

Figure 2 displays the temporal profiles of the 7 SPEs. The flare SXR fluxes are detected by GOES in the soft range of 1 -- 8 $\rm {\AA}$ (red curves), for which the profiles are shifted to the peak emission times as marked by the red vertical line. The time derivatives of SXR fluxes are generally considered as a good proxy of the non-thermal HXR fluxes (Neupert 1968), which are indicated by the blue curves. The proton SPR times are indicated by the vertical dashed lines with an error estimate of 5 minutes. It is clear that the SPR times fall into the decay phase of the flare SXR emission, except for event 4 where the SPR time coincides with the SXR peak emission. All the SPR times of the 7 SPEs are tens of minutes later than the peak times of HXR fluxes.

   \begin{figure}
   \centering
   \includegraphics[width=12cm, angle=0]{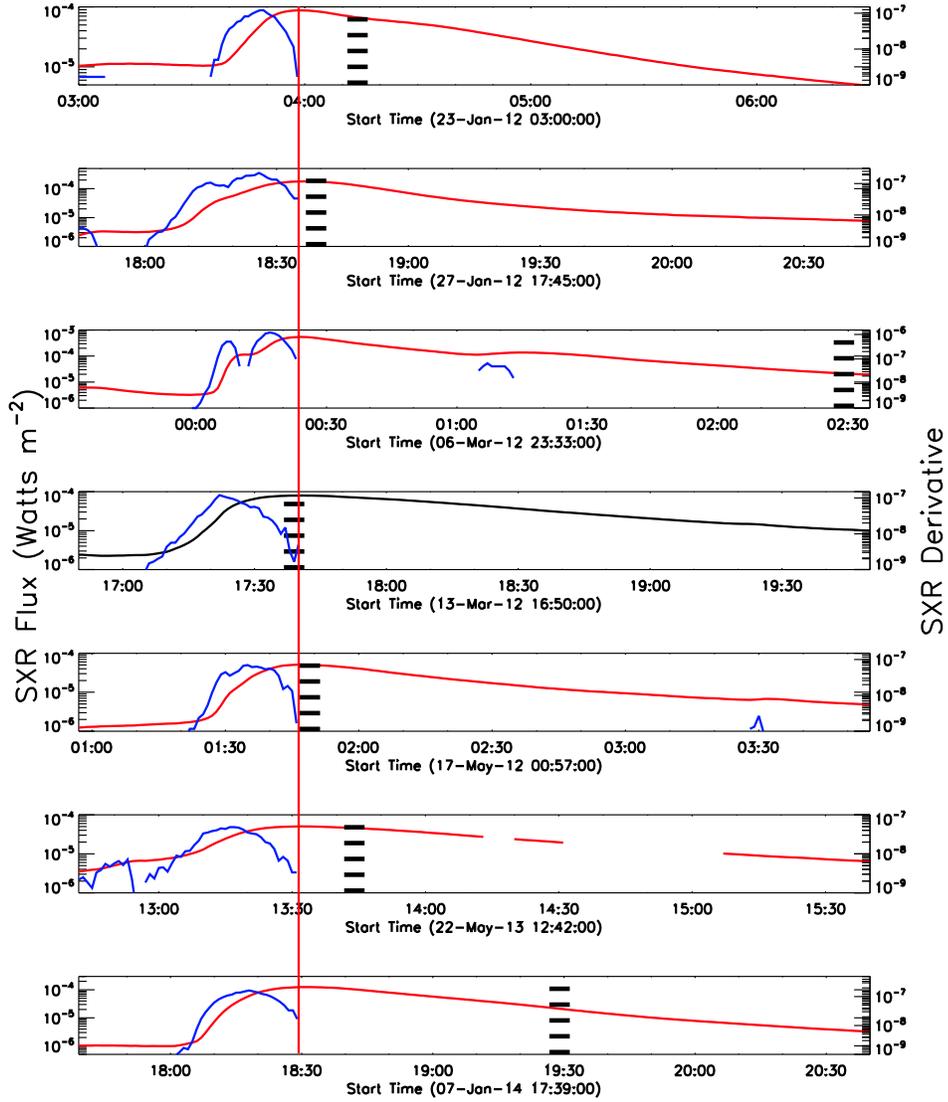}
   \caption{Temporal profiles of the 7 SPEs, in which the GOES 1 -- 8 $\rm {\AA}$ SXR fluxes are indicated with the red curves and their derivatives are with blue curves, the proton SPR times are indicated by the black dashed lines.}
   \label{Fig2}
   \end{figure}

Another interesting phenomenon is following: the farther the solar source is away from the well-connected region, the larger the time delay is between the flare SXR peak and the SPR time. For instance, the time delay of event 3 with the solar source at E15 is 125 minutes, meanwhile for the event 7 with the solar source at W11 the time delay is 58 minutes. No time delay was not observed for the event 4 which had the solar source at W62 (the well-connected region). The results seem to favor a phenomenon that the time delay is due to the stepping connection of IMF lines with the expanding CME-driven shock. However, again we can not exclude the possibility that the time delay is due to the cross-field propagation of accelerated particles from the flare site.

Under the assumption of proton acceleration by the CME-driven shock, we estimate the acceleration height derived from a linear fit of the CME speed. For event 3 and event 7, the acceleration heights are 14.3 and 12.7 solar radii ($R_{s}$) respectively, which are much larger than the other SPEs due to the poor connection of their source regions. The average acceleration height of the 7 SPEs is 7.90 $R_{s}$. To be noticed that the acceleration height of the GLE 71 is 3.07 $R_{s}$ (Li et al. 2013), which is consistent with the results of (Gopalswamy et al. 2012) and lower than the other SPEs. This suggests that the acceleration of relativistic particles occurs at lower coronal site.

\subsection{Proton spectra}

The energy spectrum of SEPs is generally considered as the indicator of their acceleration sources and mechanisms. Here we apply the method of integral time-of-maximum (TOM) spectrum, a proxy of their source spectrum (Forman et al. 1986; Miroshnichenko 1996; Miroshnichenko \& Perez-Peraza 2008). Here the proton spectra are fitted in two forms: a single power-law distribution $f(E) = AE^{-\gamma}$ and a power law broken with an exponential high-energy tail $f(E) = AE^{-\gamma}exp(-E/E_{0})$. The latter form was first obtained by Ellison $\&$ Ramaty (1985) to explain the particle acceleration by a diffusive shock. More recently, a test-particle simulation by Liu et al. (2009), with self-consistent MHD electric and magnetic fields, indicates that charged particles accelerated by the direct current (DC) electric field in the magnetic reconnection region also present a spectrum of the latter form.

   \begin{figure}
   \centering
   \includegraphics[width=12cm, angle=0]{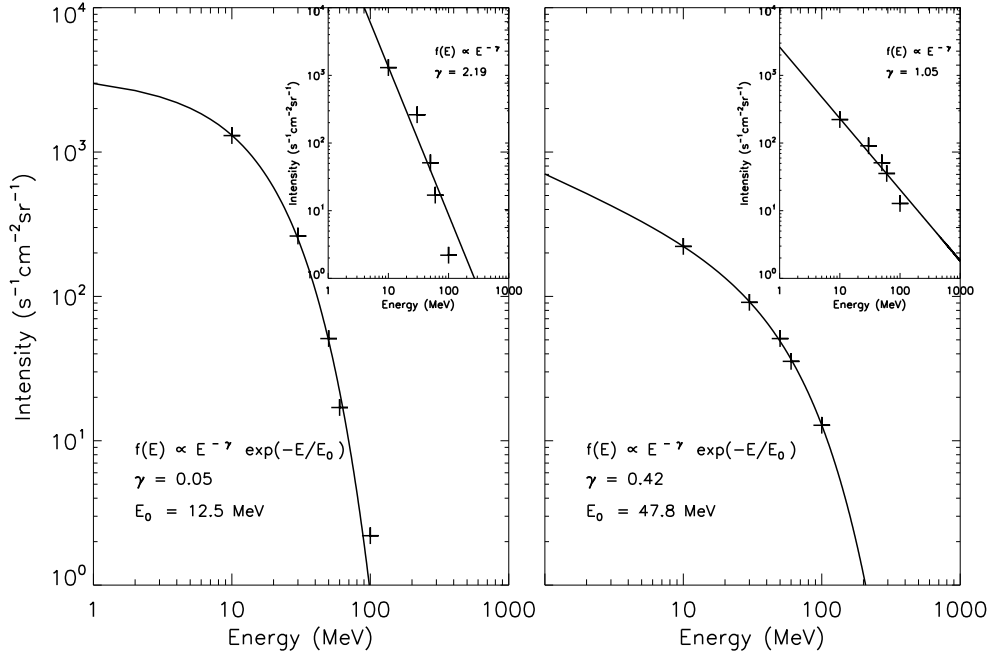}
    \vspace{0.05\textwidth}
   \caption{Integral TOM spectra of two SPEs on 2012 January 23 and 2012 May 17 (GLE71), respectively. The integral proton fluxes are obtained from the GOES 13 at the energy bands of $>$10, $>$30, $>$50, $>$60, and $>$100 MeV.}
   \label{Fig3}
   \end{figure}

\begin{deluxetable}{cccccccccccc}

\addtolength{\tabcolsep}{-4pt}
\renewcommand{\arraystretch}{1}

\tablecaption{Large Solar Proton Events of Solar Cycle 24.}

\tablehead{
\colhead{\footnotesize{Event}} & \colhead{\footnotesize{Date of}} & \colhead{\footnotesize{Intensity}} & \colhead{\footnotesize{GLE}} & \colhead{\footnotesize{Source}} &
\colhead{\footnotesize{Flare}}& \colhead{\footnotesize{CME}} & \colhead{\footnotesize{SPR}} & \colhead{\footnotesize{Flare}} & \colhead{\footnotesize{CME}} & \colhead{\footnotesize{Spectral}} & \colhead{\footnotesize{Broken}}\\

\colhead{\footnotesize{no.}} & \colhead{\footnotesize{event}} & \colhead{\footnotesize{$\geq$10 MeV}} & \colhead{\footnotesize{}} & \colhead{\footnotesize{location}} & \colhead{\footnotesize{class}} & \colhead{\footnotesize{speed}} & \colhead{\footnotesize{time}} & \colhead{\footnotesize{peak}} & \colhead{\footnotesize{height}} & \colhead{\footnotesize{index}} & \colhead{\footnotesize{energy}}\\

\colhead{} & \colhead{} & \colhead{\footnotesize{(pfu)}} & \colhead{\footnotesize{}} & \colhead{} & \colhead{} & \colhead{\footnotesize{(km/s)}} & \colhead{\footnotesize{(UT)}} &
\colhead{\footnotesize{(UT)}} & \colhead{\footnotesize{($R_{s}$)}} & \colhead{} & \colhead{\footnotesize{(MeV)}}
}

\startdata
\footnotesize{1}  & \footnotesize{2012/01/23} & \footnotesize{6310} & \footnotesize{N} & \footnotesize{N28W36} & \footnotesize{M8.7} & \footnotesize{2175} & \footnotesize{04:14} & \footnotesize{03:59} & \footnotesize{6.59} & \footnotesize{2.19} & \footnotesize{12.5} \\
\footnotesize{2}  & \footnotesize{2012/01/27} & \footnotesize{796}  & \footnotesize{N} & \footnotesize{N27W71} & \footnotesize{X1.7} & \footnotesize{2508} & \footnotesize{18:39} & \footnotesize{18:36} & \footnotesize{7.02} & \footnotesize{1.85} & \footnotesize{17.5} \\
\footnotesize{3}  & \footnotesize{2012/03/07}    & \footnotesize{6530} & \footnotesize{N} & \footnotesize{N17E15} & \footnotesize{X5.4} & \footnotesize{1825} & \footnotesize{02:29} & \footnotesize{00:24} & \footnotesize{14.3} & \footnotesize{1.25} & \footnotesize{35.7} \\
\footnotesize{4}  & \footnotesize{2012/03/13}   & \footnotesize{469}  & \footnotesize{N} & \footnotesize{N18W62} & \footnotesize{M7.9} & \footnotesize{1884} & \footnotesize{17:39} & \footnotesize{17:40} & \footnotesize{4.71} & \footnotesize{1.39} & \footnotesize{19.5} \\
\footnotesize{\textbf{5}} & \footnotesize{\textbf{2012/05/17}} & \footnotesize{\textbf{255}} & \footnotesize{\textbf{Y}} & \footnotesize{\textbf{N12W83}} & \footnotesize{\textbf{M5.1}}
& \footnotesize{\textbf{1582}} & \footnotesize{\textbf{01:49}} & \footnotesize{\textbf{01:47}} & \footnotesize{\textbf{3.07}} & \footnotesize{\textbf{1.05}} &\footnotesize{\textbf{47.8}} \\
\footnotesize{6}  & \footnotesize{2013/05/22}     & \footnotesize{1660} & \footnotesize{N} & \footnotesize{N15W70} & \footnotesize{M5.0} & \footnotesize{1466} & \footnotesize{13:44} & \footnotesize{13:32} & \footnotesize{6.91} & \footnotesize{1.36} & \footnotesize{35.0} \\
\footnotesize{7}  & \footnotesize{2014/01/07}  & \footnotesize{1033} & \footnotesize{N} & \footnotesize{S15W11} & \footnotesize{X1.2} & \footnotesize{2095} & \footnotesize{19:29} & \footnotesize{18:31} & \footnotesize{12.7} & \footnotesize{1.70} & \footnotesize{16.6}
\enddata

\footnotesize{\tablecomments{Bold text indicates the parameters of GLE71. The spectral index is derived from a single power-law distribution, and the broken energy is from the power law roll over with an exponential tail.}}

\end{deluxetable}

Figure 3 shows the integral TOM spectra of two SPEs on 2012 January 23 and 2012 May 17 (GLE71), respectively. For the power-law distribution, the spectral index is 2.19 for 2012 January 23 event, that is much softer than that for the 2012 May 17 GLE event ($\sim$1.05). For the distribution of power law broken with an exponential tail, the broken or roll-over energy of 2012 January 23 event is 12.5 MeV, that is much smaller than the 2012 May 17 GLE event with the roll over energy being 47.8 MeV. The spectral index and roll over energy of the 7 SPEs are listed below in Table 1. It is clear that, even though, the GLE 71 displays a lowest proton peak flux of 255 pfu at energy threshold $>$10 MeV, it demonstrates, however, a hardest power law spectrum, and the broken energy for the exponential high-energy tail is larger than in the other SPEs. This explains why the unique GLE event can extend to relativistic energy.

Of special interest is that a recent SPE on 2014 January 6 (one day before the event 7) was recorded with identified effects at some ground-based cosmic ray stations, for instance, at two NMs at the South Pole (SOPO and SOPB), according to the data of the Neutron Monitor Data Base (NMDB, $http://www.nmdb.eu$). The peak flux of the SPE is only 40 pfu at energy threshold $>$10 MeV. We have also derived the integral TOM spectrum of the SPE or the ``hidden GLE72". The spectral index is 0.33 for the power law distribution, and the roll over energy is 103.6 MeV for the exponential high-energy tail. Therefore, the ``hidden GLE72" might have a certain amount of relativistic particles that give rise to the response of the polar cosmic ray stations.

\section{Summary and discussion}

We have investigated 7 most intensive SPEs of the current solar cycle 24 up to date, including the unique GLE71 on 2012 May 17. Table lists a summary of the SPEs and the associated flares and CMEs. The in situ proton measurements and remote sensing solar observations lead to the following results: (1) The initial time-intensity profiles are highly dependent of the locations of solar sources, with gradual increases from active regions located in the eastern hemisphere and very impulsive increases from sources in the well connected regions. (2) The SPR times are in the decay phase of the flare emission, and the time delays between the SPRs and SXR peaks are also dependent of the locations of solar sources. (3) The proton acceleration occurs when the CME travels to 3 -- 15 $R_{s}$ with an average height of 7.9 $R_{s}$. The acceleration height of the GLE71 is 3.07 $R_{s}$, suggesting relativistic particles are accelerated in a relatively lower coronal site. (4) Comparing the other SPEs, the GLE71 displays a hardest power law spectrum and a largest broken energy for the exponential high-energy tail. This explains why the unique GLE event can extend to relativistic energy.

Even though the observational results appear to support the concept of proton acceleration by the CME-driven shock, we can not exclude the possibility of particle acceleration at flare site with subsequent perpendicular diffusion or cross-field propagation in IMF, as mentioned before. On the other hand, as demonstrated by Aschwanden (2012) extended acceleration in the flare site is possible and can be diagnosed by prolonged HXR and gamma-ray emission in many GLE events. Therefore, the time delay between the SPR and flare peak emission can be explained by extended particle acceleration and/or trapping.

Another fact comes from this study is that strong flares and fast CMEs are not the necessary requisites for the GLE event or relativistic particles, and the flux of a SPE is not the indicator for relativistic particles. Particles can be accelerated to relativistic ranges under certain conditions, coronal magnetic topologies, electric fields in the reconnection regions, plasma parameters in solar corona, compression ratios and geometries of shocks, etc.

It is timely to mention here a preliminary result of the analysis of the CARPET cosmic ray detector (Brasilia) records on 2011 March 7 and 2012 January 23 (Makhmutov et al. 2013). These authors conclude: (1) statistically significant increases were detected during 20:10 - 21:30UT on 2011 March 7 and during 03:30 - 08:00 UT on 2012 January 23; (2) these increases are indications of the long-lasting presence of high-energy solar protons ($>$9 GeV). Independent results of the analysis of the characteristics of VLF propagation and riometer records during these events support this conclusion. They note that a more careful analysis of the NMDB records is needed in order to get a final conclusion on the presence or not of solar flare effects in the form of GLEs during these events. Our analysis supports these conclusions and reflects the necessity to measure low-intensity fluxes of SCRs (see also Miroshnichenko et al. 2013).

\begin{acknowledgements}
We are grateful to the GOES and NMDB for providing observational data in this study. This work is supported by the project 985 of Nanjing University and Advanced Discipline Construction Project of Jiangsu Province and NKBRSF under grants 2014CB744203. C. Li would like to thank the Natural Science Foundation (BK2012299) of Jiangsu province and NSFC under grants 11303017. Work by L.I.M. is partially supported by the Russian Foundation of Basic Research (projects 13-02-00612 and 13-02-91165).
\end{acknowledgements}

\end{document}